\documentclass[%
 aip,
 amsmath,amssymb,
 reprint,%
]{revtex4-2}

\usepackage{graphicx}
\usepackage{blindtext}
\usepackage{color}
\usepackage{dcolumn}

\begin{document}

\title{Flux ramp modulation based hybrid microwave SQUID multiplexer}

\author{Constantin Schuster}
\affiliation{Kirchhoff-Institute for Physics, Heidelberg University, Im Neuenheimer Feld 227, 69120 Heidelberg, Germany.}
\affiliation{Institute of Micro- and Nanoelectronic Systems, Karlsruhe Institute of Technology, Hertzstraße 16, 76187 Karlsruhe, Germany.}
\author{Mathias Wegner}
\affiliation{Kirchhoff-Institute for Physics, Heidelberg University, Im Neuenheimer Feld 227, 69120 Heidelberg, Germany.}
\affiliation{Institute of Micro- and Nanoelectronic Systems, Karlsruhe Institute of Technology, Hertzstraße 16, 76187 Karlsruhe, Germany.}
\author{Christian~Enss}
\affiliation{Kirchhoff-Institute for Physics, Heidelberg University, Im Neuenheimer Feld 227, 69120 Heidelberg, Germany.}
\author{Sebastian~Kempf}
\email[]{sebastian.kempf@kit.edu}
\affiliation{Kirchhoff-Institute for Physics, Heidelberg University, Im Neuenheimer Feld 227, 69120 Heidelberg, Germany.}
\affiliation{Institute of Micro- and Nanoelectronic Systems, Karlsruhe Institute of Technology, Hertzstraße 16, 76187 Karlsruhe, Germany.}

\date{\today}

\begin{abstract}
We present a hybrid microwave SQUID multiplexer that combines two frequency-division multiplexing techniques to allow multiplexing a given number of cryogenic detectors with only a fraction of frequency encoding resonators. Similar to conventional microwave SQUID multiplexing, our multiplexer relies on inductively coupling non-hysteretic, unshunted rf-SQUIDs to superconducting microwave resonators as well as applying flux ramp modulation for output signal linearization. However, instead of utilizing one resonator per SQUID, we couple multiple SQUIDs to a common readout resonator and encode the SQUID input signals in sidebands of the microwave carrier by varying the flux ramp modulation frequency for each SQUID. We prove the suitability of our approach using a prototype device and argue by means of fundamental information theory that our approach is particularly suited for reading out large cryogenic bolometer arrays.
\end{abstract}


\maketitle

Cryogenic detectors such as superconducting transition edge sensors \cite{Irw05, Ull15} (TESs) or magnetic microcalorimeters \cite{Fle05, Kem18} (MMCs) are among the most sensitive devices for measuring incident power or radiation. Using an ultra-sensitive thermometer, either based on superconducting (TESs) or paramagnetic (MMCs) materials, as well as an appropriate readout circuit, they convert the input signal into a change of current or magnetic flux that can be precisely measured using wideband superconducting quantum interference devices (SQUIDs)\cite{Fag06}. While the maturity of fabrication technology allows building detector arrays of virtually any size, the lack of suitable SQUID based multiplexing techniques presently somehow limits the number of detectors that can be used in practice.

Existing SQUID multiplexers rely on time-division \cite{Dor16}, frequency-division using MHz \cite{Har14, Ric21} or GHz carriers \cite{Mat08, Hir13, Kem17}, code-division \cite{Mor16} or hybrid \cite{Rei08, Irw18, Yu20} multiplexing schemes.  Among those, microwave SQUID multiplexers \cite{Mat08, Hir13, Kem17} ($\mu$MUXs) are the devices of choice when it comes to read out ultra-large scale detector arrays employing tens or hundreds of thousands of individual detectors. For a $\mu$MUX, each readout channel comprises a non-hysteretic, unshunted rf-SQUID \cite{Han73} which transduces the detector signal into a change of amplitude or phase of a microwave signal probing a superconducting microwave resonator that is inductively coupled to the SQUID. For actual multiplexing, many readout channels, each comprising a microwave resonator with unique resonance frequency, are capacitively coupled to a common transmission line (feedline). This allows for simultaneously monitoring the state of all detectors by injecting a frequency comb and continuously measuring the amplitude or phase of each carrier. 

The channel capacity of the transmission line ultimately limits the number of detectors than can be simultaneously read out using a single $\mu$MUX \cite{Sha49}. In practice, however, there are several effects reducing the maximum channel count. Most importantly, cryogenic amplifiers used for boosting the multiplexer output signal limit the usable frequency range and set the noise level of the overall system. Saturation power and third-order intercept point of these amplifiers further reduce the channel capacity \cite{Sat20}. Moreover, the signal rise time sets the response time and hence the bandwidth of the resonators and thus limits their density in frequency space. The latter results from the necessity of spacing the resonators far enough to minimize inter-resonator crosstalk due to the overlap of Lorentzian resonance tails \cite{Mat19}. Hybrid multiplexing techniques might allow to tackle resulting challenges and potentially even allow to use the channel capacity of the transmission line more efficiently. As an alternative to existing hybrid techniques \cite{Rei08, Irw18, Yu20}, we present a hybrid multiplexing technique combining conventional microwave SQUID multiplexing and flux ramp modulation based cryogenic SQUID multiplexing \cite{Ric21}.

\begin{figure}
    \includegraphics[width=1.0\linewidth]{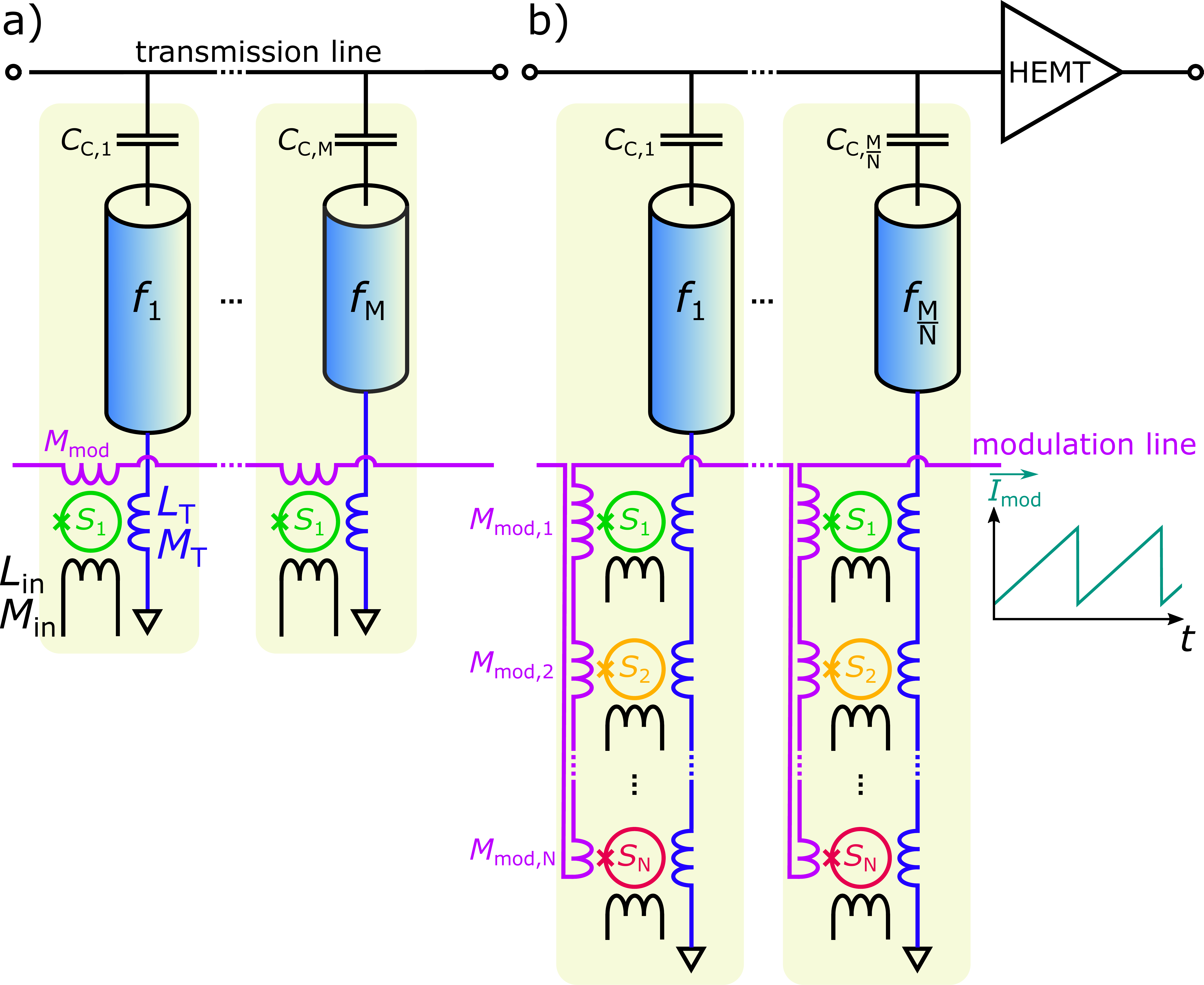}%
    \caption{\label{fig1}(Color online)
    (a) Schematic circuit diagram of a conventional microwave SQUID multiplexer ($\mu$MUX). It comprises $M$ readout resonators that are each capacitively coupled to a common transmission line and inductively coupled to a non-hysteretic, current-sensing rf-SQUID to read out $M$ input signals. For output signal linearization by means of flux ramp modulation, a modulation coil is inductively coupled to each SQUID. (b) Schematic circuit diagram of a hybrid microwave SQUID multiplexer (H$\mu$MUX). It comprises $M/N$ readout resonators to read out $M$ input signals. For this, each readout resonator is inductively coupled to $N$ non-hysteretic, current-sensing rf-SQUIDs modulating together the resonance frequency. For distinguishing the individual signals and output signal linearization, a flux ramp modulation coil is coupled to the individual SQUIDs of a readout resonator with different strength to yield sideband modulation.}
\end{figure}

For conventional $\mu$MUXs (see Fig.~\ref{fig1}a), flux ramp modulation \cite{Mat12} causes a sawthooth shaped variation of the magnetic flux threading the SQUID loops and thus a periodic modulation of the SQUID inductances. The effective modulation frequency $f_{\mathrm{mod}} = I_\mathrm{mod} M_{\mathrm{mod}} f_\mathrm{ramp} / \Phi_0$ depends on the slope $I_\mathrm{mod} f_\mathrm{ramp}$ of the flux ramp as well as the mutual inductance $M_\mathrm{mod}$ between the SQUID and the modulation coil. Here, $f_\mathrm{ramp}$ denotes the ramp repetition rate. It must be chosen such that an external signal is quasi-static within the period of the flux ramp. An external flux signal hence causes a quasi-static phase shift of the SQUID response that is proportional to the input signal.

In contrast, our hybrid microwave SQUID multiplexer (H$\mu$MUX, see Fig.~\ref{fig1}b) relies on inductively coupling $N$ individual non-hysteretic, unshunted rf-SQUIDs, each connected to an individual detector, to a single resonator. To encode the signal information of the $N$ SQUIDs in sidebands of the microwave carrier, we vary the modulation frequency $f_{\mathrm{mod,}i}$ for each SQUID. For this, we wire all modulation coils in series and inject a periodic, sawtooth-shaped current signal with amplitude $I_\mathrm{mod}$ and repetition rate $f_\mathrm{ramp}$ into the common modulation coil. As we systematically vary the mutual inductance $M_{\mathrm{mod,}i}$ between modulation coil and SQUID loop (see discussion below), the modulation frequencies $f_{\mathrm{mod,}i}$ are unique. For this reason, superimposing the responses of all SQUIDs onto a single microwave carrier allows reconstructing the individual signals by means of demodulation of the multiplexer output data stream with the known modulation frequencies $f_{\mathrm{mod,}i}$. With this H$\mu$MUX approach, each microwave resonator hence reads out $N$ rf-SQUIDs in parallel. Compared to conventional $\mu$MUXs, this reduces the number of resonators and probe tones required for reading out a fixed number of detector channels by a factor of $N$.

\begin{figure*}
    \includegraphics[width=1.0\linewidth]{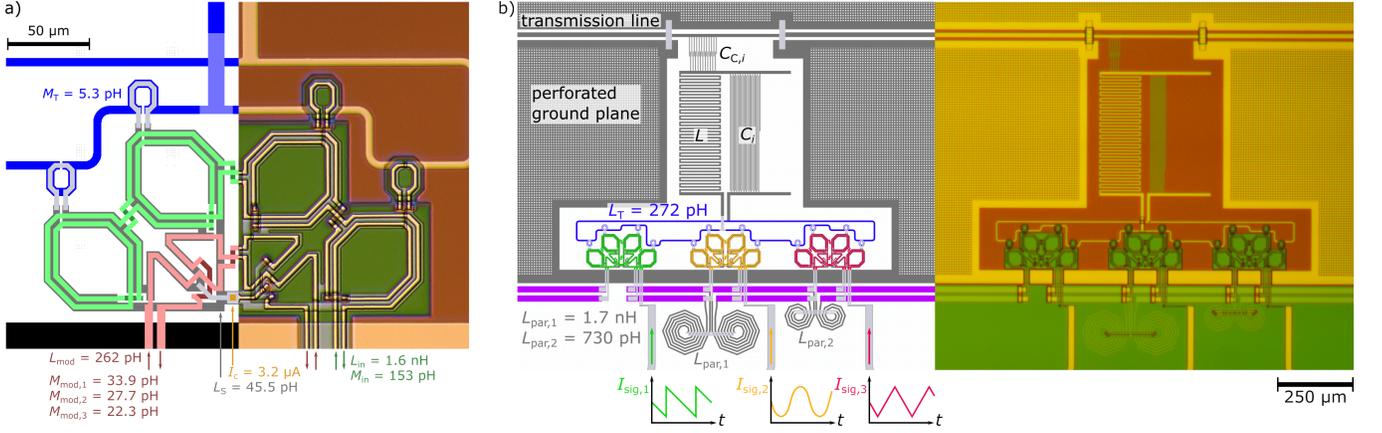}%
    \caption{\label{fig2}(Color online)
    (a) Composite image of the layout (left) and a microscope picture (right) of one of the non-hysteretic rf-SQUIDs that are used as parametric inductor to modulate the resonance frequency of the individual microwave resonators. The SQUID is formed by four superconducting loops that are connected in parallel. On top or below of each loop, the load inductor (blue), the input coil (green) and the modulation coil (pink) are running. The notation of the different inductance refers to Fig.~\ref{fig1}.
    (b) Composite image of the layout (left) and a microscope picture (right) of two neighboring H$\mu$MUX resonator channels. The resonators are formed by lumped element Nb based resonators. The resonance frequency is set by varying the length and number of the fingers of the interdigital capacitor. The load inductor (blue) is inductively coupled to three independent current-sensing SQUIDs (green, yellow, red) to read out three signals with only one readout resonator. For varying the mutual inductance $M_\mathrm{mod,i}$ between the different SQUIDs and the load inductor, gradiometric spiral inductors of different sizes are connected in parallel to the modulation coil of each SQUID. In the layout picture, the input signals for validating our readout concept are shown.}
\end{figure*}

For proving the suitability of our approach, we designed, fabricated and characterized a non-optimized H$\mu$MUX prototype that is based on lumped-element microwave resonators, each being formed by a meander-shaped inductor $L$, an interdigital capacitor $C$ as well as a load inductor $L_\mathrm{T}$ connected in series (see Fig.~\ref{fig2}). The latter is geometrically shaped such that it evenly couples to $N =3$ independent non-hysteretic rf-SQUIDs via the mutual inductance $M_\mathrm{T}$. Using an interdigital coupling capacitor $C_\mathrm{C}$, each resonator is coupled to a common coplanar waveguide (CPW) transmission line allowing the carrier signals sent from a dedicated room-temperature readout electronics to pass by the resonators. The CPW ground plane covers a big fraction of the chip surface and is perforated with $4\,\mathrm{\mu m} \times 4\,\mathrm{\mu m}$ square holes to avoid motion of trapped vortices within the ground plane potentially degrading the internal quality factor of the microwave resonators.

Each rf-SQUID forms a gradiometer to minimize the influence of external disturbances emerging from background magnetic fields and is coupled to in total three coils, i.e. the load inductor $L_\mathrm{T}$ for coupling the SQUID to the resonator, the input coil $L_\mathrm{in}$ to couple the  signal into the SQUID loop, and the common flux ramp modulation coil $L_\mathrm{mod}$. For varying the mutual inductances $M_\mathrm{mod}$ among the SQUIDs that are coupled to the same microwave resonator, we connect locally superconducting inductances $L_\mathrm{par,i}$ in parallel to the modulation coil (see Fig.~\ref{fig2}b). In comparison to varying the geometric overlap between the modulation coil and the SQUID loop as we have used for our flux ramp modulation based MHz FDM SQUID multiplexer\cite{Ric21}, this approach is beneficial as it allows for fine-trimming the mutual inductances in a post-processing step to guarantee matching the boundary conditions for the modulation frequencies (see discussion below).

We mounted the fabricated prototype multiplexer on a custom-made sample holder, connected the chip electrically to a custom printed circuit board using ultrasonic wedge bonds and immersed the setup into a liquid helium transport dewar. Due to space constraints within the measuring dipstick, we couldn't use a cryogenic HEMT amplifier for boosting the multiplexer output signal. We used instead two room-temperature microwave amplifiers. For basic device characterization, we used a vector network analyzer in combination with an arbitrary waveform generator injecting a static flux or flux ramp signal into the modulation line. For multiplexing demonstration, we used our custom SDR based readout electronics \cite{Weg18, San19} to continuously stream the output signal of the multiplexer. Demodulation of the modulated signal was then performed offline.

\begin{figure}[t]
    \includegraphics[width=1.0\linewidth]{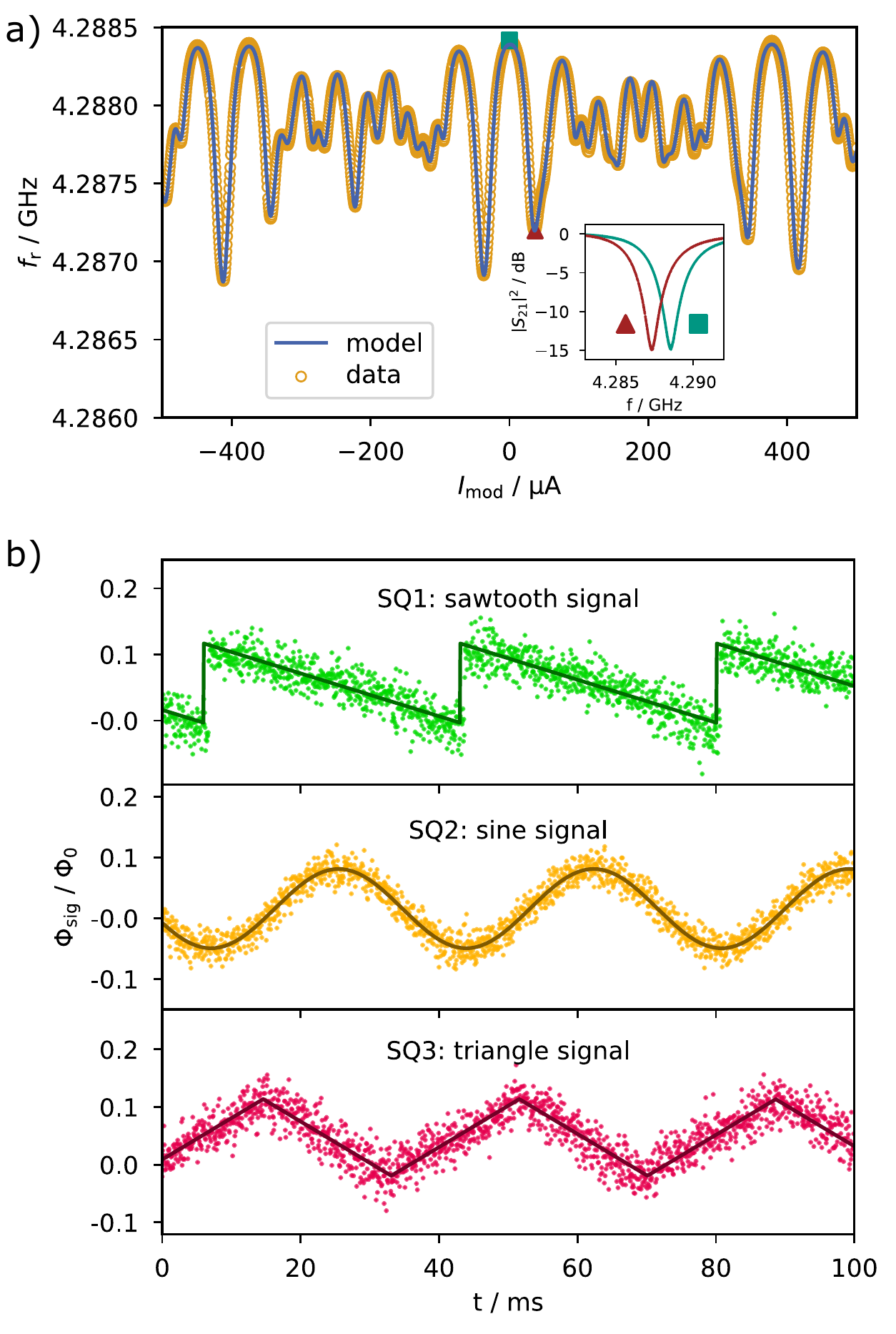}%
    \caption{\label{fig3}(Color online) (a) Dependence of the measured resonance frequency of one H$\mu$MUX resonator readout channel on a dc-current driven through the current modulation coil. The solid line depicts the prediction of the course as yielded by an adaption of our most recent analytical $\mu$MUX model \cite{Weg21}. The inset shows two exemplary resonance curves for points in the plot marked by a red triangle or a green square.
    (b) Measured magnetic flux $\Phi_\mathrm{sig}$ injected in SQUID ``SQ$i$'' versus time $t$ as derived from the demodulation of the output signal of one readout resonator of our H$\mu$MUX. The test signals that have been applied to the input coils of the different SQUIDs are illustrated in Fig.~\ref{fig2}(b). The solid lines show the injected input signals.}
\end{figure}

Fig.~\ref{fig3}a shows the measured dependence of the resonance frequency $f_\mathrm{r}(I_\mathrm{mod})$ of an exemplary resonator on the strength of a dc current through the modulation coil as well as the expected behavior that was calculated using an adaption of our most recent analytical model of a microwave SQUID multiplexer \cite{Weg21}. The agreement is excellent and shows that the device behavior can be reliably predicted. Moreover, Fig.~\ref{fig3}b depicts a time section of three demodulated output data stream when injecting three different signals into the input coil of the SQUIDs. The input signals are clearly resolved, hence proving that our multiplexing approach is working as intended. Although the noise level of the present measurement is rather high due to the missing cryogenic HEMT amplifier, the perfect agreement with our multiplexer model let us expect that noise levels close to that of conventional $\mu$MUXs can be achieved. However, we have to expect a noise penalty as we will explain in the following paragraphs.

The signal bandwidth per readout channel, the multiplexing factor, i.e. the number of readout channels that can be simultaneously read out using a single transmission line, as well as the achievable noise level are three key figures of merit that benchmark any cryogenic multiplexer. To check for these benchmarks, we consider a single readout resonator to which $N$ non-hysteretic rf-SQUIDs are coupled. To minimize crosstalk, we must ensure that different carrier frequencies aren't multiples of each other. Since the ramp repetition rate $f_\mathrm{ramp}$ is identical for all SQUIDs, the modulation frequency $f_{\mathrm{mod,}i} = A_{\mathrm{mod,}i} f_\mathrm{ramp}$ can only be varied by changing the modulation amplitude $A_{\mathrm{mod,}i}$ according to
    \begin{equation}
        A_{\mathrm{mod,}1} < A_{\mathrm{mod,}2} < \ldots < A_{\mathrm{mod,}N} < 2 A_{\mathrm{mod,}1}
    \end{equation}
with $A_{\mathrm{mod,}i} \in \mathbb{N}$. For simplicity, we assume that each ramp segment is fully used for phase reconstruction, i.e. we neglect that in real applications the data stream is typically truncated to remove transients etc \cite{Mat12}. The modulation amplitudes must hence be integers for the demodulation algorithm to work reliably. As the signal sampling rate is identical to the ramp repetition rate $f_\mathrm{ramp}$, the latter should be as high as possible. Moreover, the effective modulation frequencies $f_{\mathrm{mod,}i}$ must  be smaller than a limit frequency $f_\mathrm{lim}$ existing due to the finite resonator response time, thus requiring the modulation amplitudes to be minimized. To ensure both, unique and integer-valued modulation amplitudes $A_\mathrm{mod,i+1} - A_{\mathrm{mod,}i} \geqslant 1$, the smallest difference between them is unity. If we choose the minimum value as the difference between smallest and highest modulation amplitudes, we hence yield
    \begin{eqnarray}
        A_{\mathrm{mod,}N}^\mathrm{min} + 1 \overset{!}{=} 2 A_\mathrm{mod,1}^\mathrm{min}
         & \rightarrow & A_\mathrm{mod,1}^\mathrm{min} + N 
          \overset{!}{=}  2 A_\mathrm{mod,1}^\mathrm{min} \nonumber\\
         & \rightarrow & A_\mathrm{mod,1}^\mathrm{min}  \overset{!}{=}  N.
    \end{eqnarray}
The signal bandwidth is limited by the modulation frequency of the SQUID with the largest modulation amplitude for which we yield according to the previous discussion the relation $f_\mathrm{mod}^\mathrm{max} = f_{\mathrm{mod,}N} = \left( 2N - 1 \right) f_\mathrm{ramp}$.
 
We usually want to maximize $f_\mathrm{ramp}$ to yield a high sampling rate of the signal. We therefore fix $f_\mathrm{mod}^\mathrm{max} = f_\mathrm{lim} \propto \Delta f_\mathrm{BW}$. Here, $\Delta f_\mathrm{BW}$ denotes the bandwidth of the readout resonator. We introduce the density of SQUIDs in frequency space $\rho_\mathrm{SQ} = N_\mathrm{SQ} / \Delta f_\mathrm{tot} = N / \left( \Delta n_\mathrm{opt} \Delta f_\mathrm{BW} \right)$, where $\Delta f_\mathrm{tot}$ is the full bandwidth available that is usually limited by cryogenic amplifiers used to boost the tiny multiplexer output signals. Moreover, $N_\mathrm{SQ}$ is the total number of SQUIDs of all multiplexer channels that can be read out using a single feedline and $\Delta n_\mathrm{opt}$ denotes a guard factor ensuring that the spacing between neighboring resonators is sufficiently large to minimize interchannel crosstalk. This yields the result
    \begin{equation}
        \left( 2N - 1 \right) f_\mathrm{ramp} = f_\mathrm{lim} \propto \Delta f_\mathrm{BW} \propto \frac{N}{\rho_\mathrm{SQ}}.
    \end{equation}
from which we immediately conclude:
    \begin{equation}\label{penalty}
        f_\mathrm{ramp} \rho_\mathrm{SQ} \propto \frac{N}{\left( 2N - 1 \right)} = 
        \begin{cases}
            1 \ &\mathrm{for}\ N = 1\\
            \frac{1}{2}\ &\mathrm{for}\ N \rightarrow \infty
        \end{cases}
    \end {equation}
We aim for maximizing the product $f_\mathrm{ramp}\rho_\mathrm{SQ}$ to yield a large density of SQUIDs as well as a high signal sampling rate. From Eq.~\eqref{penalty} we see that the maximum is reached for $N = 1$, i.e. the channel capacity of the feedline is most effectively used for conventional microwave SQUID multiplexing. The largest penalty, i.e. $f_\mathrm{ramp} \rho_\mathrm{SQ} = 1/2$, occurs in the limit of $N \rightarrow \infty $, for which an infinitely large amount of SQUIDs is coupled to a single readout resonator. For the hybrid microwave SQUID multiplexer, the necessity for distinct modulation frequencies between SQUIDs coupled to the same resonator prohibits the modulation with the minimum amplitude for all SQUIDs, leading to a less efficient use of the available bandwidth.

In addition to the bandwidth penalty, hybrid microwave SQUID multiplexing possesses another disadvantage with respect to noise caused by the cryogenic amplifier chain. The flux-to-voltage transfer coefficient
    \begin{equation}
        V_\mathrm{\Phi}^{\mathrm{max,}i} \propto \frac{\Delta f_\mathrm{res}^{\mathrm{max,}i}}{\Delta f_\mathrm{BW}} \sqrt{P_\mathrm{exc}}
    \end{equation}
determines the effective flux noise in the $i$-th SQUID that is caused by the cryogenic amplifier chain. Here, $\Delta f_\mathrm{res}^{\mathrm{max,}i}$ denotes the maximum resonance frequency shift caused by the $i$-th SQUID and $P_\mathrm{exc}$ is the power of the resonator probing signal. Flux ramp modulation works best if the resonator bandwidth $\Delta f_\mathrm{BW}$ equals the total resonance frequency modulation amplitude $\Delta f_\mathrm{res}^\mathrm{max,tot} = \sum \Delta f_\mathrm{res}^{\mathrm{max,}i}$. Assuming that all SQUIDs have the same modulation amplitude $\Delta f_\mathrm{res}^{\mathrm{max,}i}$, we hence yield
    \begin{equation}
        V_\mathrm{\Phi}^{\mathrm{max,}i} \propto  \frac{\sqrt{P_\mathrm{exc}}}{N}.
    \end{equation}
For applications requiring a large number of readout channels, the total readout power $P_\mathrm{exc,tot} = N_\mathrm{res} P_\mathrm{exc}$ is limited by the saturation power of the HEMT amplifier rather than the optimal readout power for each channel. With hybrid SQUID multiplexing, the number of resonators $N_\mathrm{res}$ required to read out a fixed number of SQUIDs is decreased by a factor of $N$. If the overall readout power $P_\mathrm{exc,tot}$ is kept constant, this allows increasing the readout power per channel $P_\mathrm{exc}$ accordingly. For this reason, the total penalty to the flux noise per channel caused by the HEMT is given by
    \begin{equation}\label{eq:hymux_noise_03}
        S_\mathrm{\Phi,HEMT,i} = N\tilde{S}_\mathrm{\Phi,HEMT}
    \end{equation}
Here, 
if $\tilde{S}_\mathrm{\Phi,HEMT}$ is apparent HEMT related flux noise of a single readout channel of a conventional microwave SQUID multiplexer with the corresponding set of device- and readout parameters.

Our simple discussion predicts a penalty in both, the maximum readout channel density as well as the apparent flux noise of a single readout channel of a hybrid SQUID multiplexer as compared to a conventional microwave SQUID multiplexer. We would hence have to doubt whether hybrid microwave SQUID multiplexing is a useful technique for the readout of cryogenic detector arrays. However, there are strong benefits regarding the fabrication of actual devices. If a microwave SQUID multiplexer is used to read out detectors with a small signal bandwidth, e.g. for bolometric applications, the channel capacity of the feedline allows for the simultaneous readout of a tremendous number of detectors \cite{Irw09}. For a conventional microwave SQUID multiplexer, realizing a high channel count requires the fabrication of a large number of resonators which must be evenly spaced in frequency space to minimize crosstalk due to the overlap of Lorentzian resonance tails \cite{Mat19}. As the channel number increases, resonator spacing gets significantly smaller, thus requiring stringent fabrication tolerances to keep the variation of resonance frequencies small as compared to the frequency spacing. Spacing neighboring resonators with a resonance frequency variation below $1\,\mathrm{kHz}$, for example, would require to control the length of the last finger of an interdigital capacitor of a lumped-element microwave resonator (see Fig.~\ref{fig2}b) with a tolerance better than $5\,\mathrm{nm}$. This precision is quite hard to achieve using non-industrial photolithography equipment, e.g. stepper-based DUV projection immersion lithography, or time-consuming, e.g. using electron beam lithography. Moreover, tile-and-trim processes \cite{Liu17, McK19} have not yet proved to yield this ultimate level of accuracy when using conventional lithography equipment. In case that manufacturing accuracy rather than the Shannon capacity of the feedline becomes the limiting factor for the number of channels within a microwave SQUID multiplexer, hybrid microwave SQUID multiplexing hence supports to increase the multiplexing factor as a smaller number of resonators with larger frequency spacing is required to read out a fixed number of SQUIDs. As the scaling is similar to time-division multiplexing, the inherent noise penalty can even be diminished for TESs by increasing the mutual inductance between SQUID and its input coil \cite{Irw02}. For MMCs, however, the noise penalty of hybrid microwave SQUID multiplexing might remain an issue and forces to built detectors with utmost sensitivity \cite{Por13}. 

In conclusion, we have presented a hybrid microwave SQUID multiplexer combining two frequency-division multiplexing techniques to allow multiplexing a given number of cryogenic detectors or other low-impedance signal sources with only a fraction of readout resonators used for frequency encoding. It relies on the combination of conventional microwave SQUID multiplexing and flux ramp modulation based cryogenic SQUID multiplexing and provides a rather easy way to more efficiently use the channel capacity of a microwave transmission line. However, due to inherent penalties with respect to the maximum readout channel density as well as the apparent flux noise of a single readout channel, our approach turns out to most useful when fabrication accuracy becomes the limiting factor for increasing the channel count within a microwave SQUID multiplexer. For this reason, our approach is particularly suited for bolometric applications requiring a huge number of small-bandwidth cryogenic detectors.

\begin{acknowledgments}
We would like to thank T. Wolf for his great support during device fabrication and greatly acknowledge valuable discussions D. Richter, O. Sander, and N. Karcher. The work has been performed in the framework of the DFG research unit FOR 2202 (funding under grant EN299/7-2). The research leading to these results has also received funding from the European Union’s Horizon 2020 Research and Innovation Programme, under Grant Agreement No 824109.
\end{acknowledgments}

\section*{Data Availability Statement}
The data that support the findings of this study are available from the corresponding author upon reasonable request.

\bibliography{literature}

\end{document}